\documentclass[aps,prl,showpacs,twocolumn,amsmath,amssymb]{revtex4-1}
\usepackage{graphicx}
\usepackage{dcolumn}
\usepackage{bm}
\usepackage{units}
\usepackage{upgreek}
\usepackage{ucs}
\usepackage{color}
\usepackage{layouts}
\usepackage{hyperref}
\usepackage[all]{hypcap}
\preprint{APS/123-QED}

\begin{document}

\title{Cross-dimensional phase transition from an array of 1D Luttinger liquids to a 3D Bose-Einstein condensate}
\author{Andreas Vogler}
\author{Ralf Labouvie}
\author{Giovanni Barontini}
\author{Sebastian Eggert}
\author{Vera Guarrera}
\author{Herwig Ott}
\email{ott@physik.uni-kl.de}
\affiliation{Physics Dept.~and Research Center OPTIMAS, Technische Universit\"at Kaiserslautern, 67663 Kaiserslautern, Germany}

\begin{abstract}
We study the thermodynamic properties of a 2D array of coupled one-dimensional Bose gases. The system is realized with ultracold bosonic atoms loaded in the potential tubes of a two-dimensional optical lattice. For negligible coupling strength, each tube is an independent weakly interacting 1D Bose gas featuring Tomonaga Luttinger liquid behavior. By decreasing the lattice depth, we increase the coupling strength between the 1D gases and allow for the phase transition into a 3D condensate. We extract the phase diagram for such a system and compare our results with theoretical predictions. Due to the high effective mass across the periodic potential and the increased 1D interaction strength, the phase transition is shifted to large positive values of the chemical potential. Our results are prototypical to a variety of low-dimensional systems, where the coupling between the subsystems is realized in a higher spatial dimension such as coupled spin chains in magnetic insulators.
\end{abstract}

\maketitle


The emergence of new properties from low-dimensional building blocks is a universal theme in different areas in physics. In the field of material science, systems with reduced dimensions are of 
strong interest due to their peculiar properties. Prominent examples include graphene, carbon nanotubes, nanowires, and quantum dots. When such low-dimensional systems are arranged in regular patterns, the coupling between them has an additional impact: few coupled layers of graphene drastically change the thermal conductivity \cite{Gosh2010}, the transport properties of granular electronic materials can be tuned by the dimensionality \cite{Xu2009}, photonic metamaterials are governed by the cooperativity of their low-dimensional building blocks \cite{Soukoulis2011} and phase transitions can be modified by the effective dimensionality \cite{Vescoli1998}. The control of the coupling strength between the low-dimensional subsystems is essential to understand and predict the emerging properties \cite{Valla2002, Shallcross2010}. However, in real materials this control is often limited.

The investigation of transitions between isolated and coupled low dimensional systems is also at the heart of ultracold atom research. The extreme flexibility and the high degree of control has lead to the first observation of a Tonks-Girardeau gas in 1D \cite{Paredes2004, Kinoshita2004}, the superfluid to Mott insulator transition in 1D \cite{Esslinger2004}, and the BKT transition in 2D \cite{Hadzibabic2006}. The tunability of the tunnel coupling within an array of 1D or 2D quantum gases residing in an optical lattice is also ideally suited to explore the physics in the crossover between two dimensionalities. Cold atoms experiments can therefore be used to simulate real material devices where such tunability is limited. Coupled spin ladders in a magnetic field \cite{Tsvelik1999} and spin chain \cite{Affleck2002} materials constitute a paradigmatic example. In these systems, the phase transition is closely linked to the Bose-Einstein condensation of interacting bosons \cite{giamarchi2008} and can be understood in terms of a transition from an array of 1D Luttinger liquids to a 3D superfluid \cite{Ruegg2008, Thielemann2009}. The knowledge gained by means of the cold atoms simulator can thus be translated in a better control over the solid state devices.

\begin{figure}
\begin{center}
\includegraphics[width=\columnwidth]{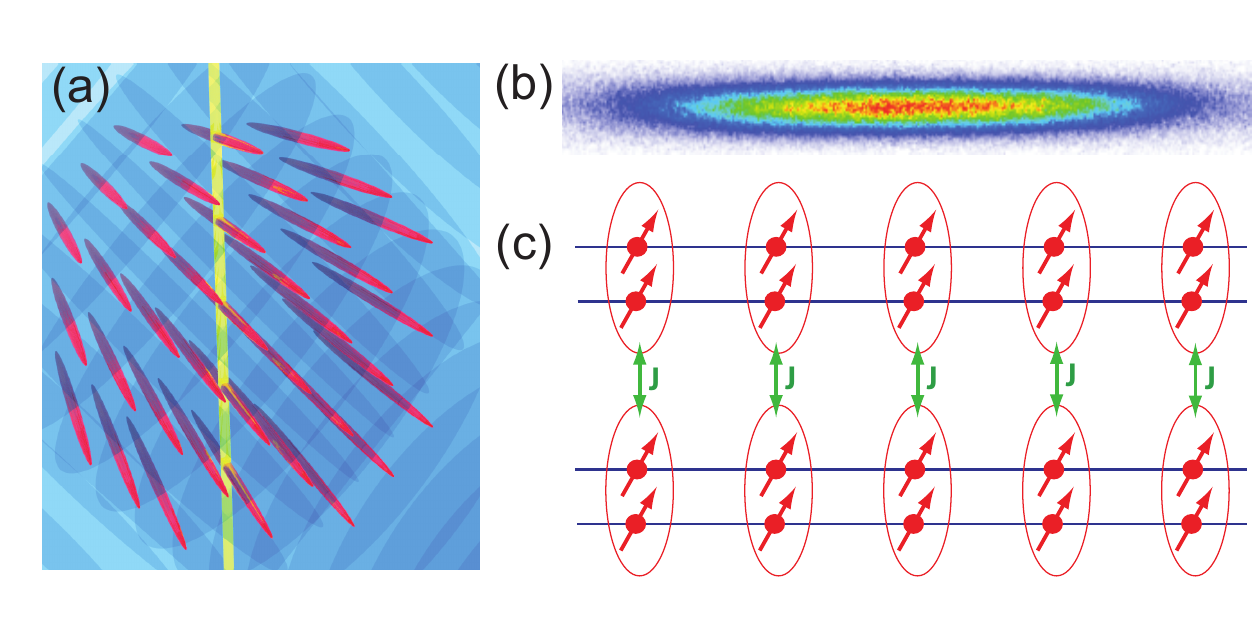}
\end{center}
\caption{(Color online). (a) Experimental setup: 1D Bose gases (red) are trapped in a 2D optical lattice (blue) with finite tunnel coupling and imaged with an electron beam (yellow) (b) \textit{In situ} image of the density distribution from which the line profiles are extracted. (c) Corresponding analogon in real materials: coupled dimer spin triplet chains show Bose-Einstein condensation above a critical magnetic field.}
\label{fig:setup}
\end{figure}

In order to realize this quantum simulator, we study ultracold bosonic rubidium atoms which are loaded in a two-dimensional optical lattice (Fig.\,1a) \cite{Fabbri2011,Vogler2013}. By tuning the depth of the optical lattice we control the coupling between the lattice sites while we use high resolution {\it in situ} imaging based on scanning electron microscopy (SEM) \cite{Gericke2008,Guarrera2012} to probe the system (Fig.\,1b). 

In the thermodynamic limit, a single 1D Bose gas does not show Bose-Einstein condensation at any temperature. For coupled 1D gases, it has been shown that a condensate phase exists for all coupling strengths at zero temperature \cite{Efetov1974}. For finite temperature, a phase transition has been predicted at a critical 1D density $n_{\rm crit}$ as a function of perpendicular coupling strength $J$ based on a mean field argument in the weak coupling limit $J\ll\mu$ \cite{Ho2004,Cazalilla2006} 

\begin{equation} 
n_{\rm crit} = f(K) J^{K/(1-2K)} T^{(4K-1)/(4K-2)}. 
\label{scaling} 
\end{equation} 

Here $1\leq K <\infty$ is the Luttinger liquid parameter which is given by $K=1$ in the Tonks limit and $K\to \infty$ for free bosons. Generally the prefactor $f(K)$ and $K$ cannot be determined analytically, but quantitative estimates can be made \cite{Cazalilla2006}. Such a phase transition is analogous to the 3D ordering of coupled spin ladder \cite{Tsvelik1999}  and spin chain \cite{Affleck2002} materials (Fig.\,1c). In a realistic setup for ultracold gases, the system is not translationally invariant. However, the inhomogeneous trapping potential can be turned into an advantage. Performing a local density approximation (LDA) \cite{Ho2009} the external potential is converted into an effective chemical potential according to $\mu_{\mathrm{eff}}(r,z)=\mu_0-V(r,z)$. Probing the system {\it in situ} with high spatial resolution therefore allows to find local indications of a phase transition, making a thermodynamic analysis possible.

In our experiment, we adiabatically load a partially condensed cloud of $7.5\times10^4$ $^{87}$Rb atoms at $T\approx 30 - 40\,$nK in a retro-reflected two-dimensional optical lattice at a wavelength of $\lambda=774\,$nm. In the lattice, the adjacent sites are coupled by the Josephson tunnel coupling $J(s)\simeq 4E_rs^{3/4}e^{-2\sqrt{s}}$ \cite{Zwerger2003}, where $s$ measures the depth of the optical lattice in units of the recoil energy $E_r=h^2/2\lambda^2 m_{\mathrm{Rb}}$. After the loading procedure we start the SEM imaging process, providing a spatial resolution of $\unit[240(10)]{nm}$. The rectangular scan pattern is oriented along the axial direction of the BEC (i.e. along the lattice tubes). As the overall trapping potential is rotationally symmetric, we deconvolve the SEM images (Fig.\,1b) by applying an inverse Abel transformation \cite{Vogler2013}. This yields the 3D density $n_\mathrm {3D}(r,z)$, which is converted into an effective 1D density (and vice versa) by multiplication with the transverse extension of a tube: $n_{\mathrm{1D}}(r,z)=(\lambda/2)^2 n_\mathrm {3D}(r,z)$. Throughout this work, all densities are given as 1D densities. The temperature $T$ and the central chemical potential $\mu_0$ are determined by Gaussian fits to the thermal wings \cite{Ho2009} and by comparison with the exact 1D thermodynamic theory \cite{Yang1969,Kheruntsyan2005}. In addition to the density profile we perform standard time of flight (TOF) absorption imaging.

\begin{figure}
\begin{center}
\includegraphics[width=1\columnwidth]{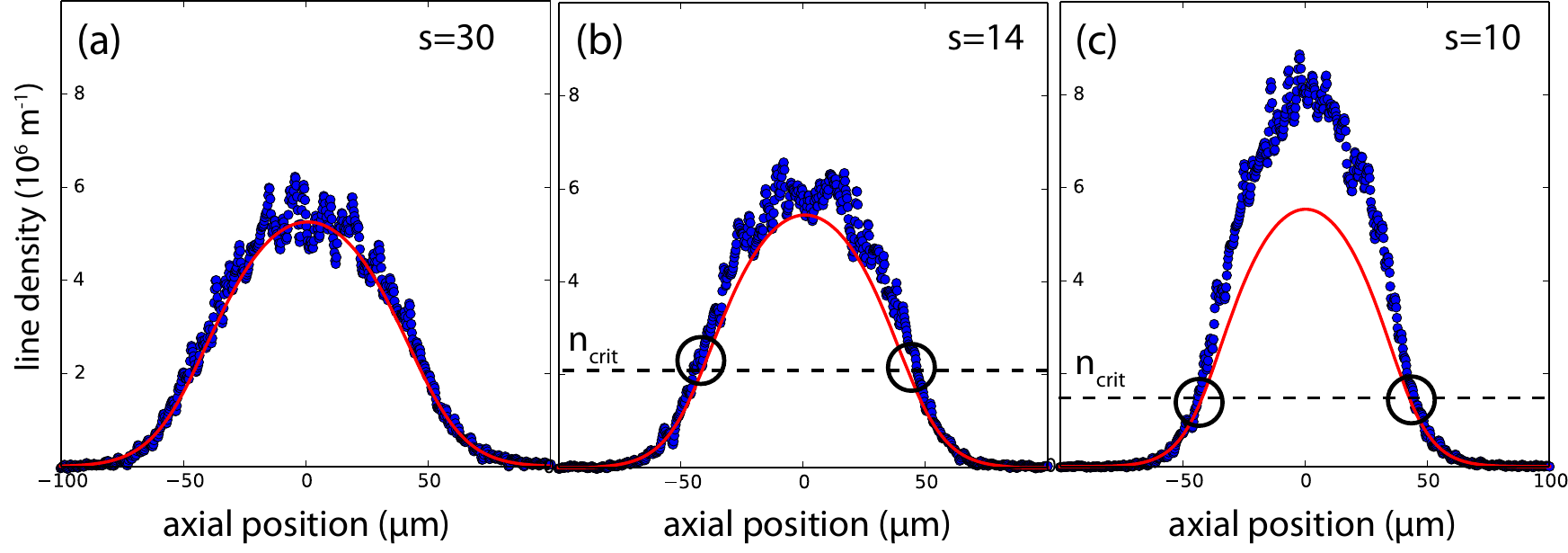}
\end{center}	
\caption{(Color online). Density profiles (blue points) for different lattice depths ((a) s=30, (b) s=14, (c) s=10, at a distance of 1.2\,$\upmu$m from the trap center. The red line is a fit with the 1D theory, where only the outer parts of the profile ($\gamma>2$) were included in the fit. The circles indicate the position, where the experimental profiles deviate from the pure 1D behaviour by more than one standard deviation.}
\label{fig:cfrac}
\end{figure}

We start the analysis with deep optical lattices ($s>20$), where the tunnel coupling between different tubes is negligible. In this limit, our experimental system consists of an array of independent 1D gases. The corresponding TOF images show no interference pattern, signaling that no phase coherence between the tubes is present. The measured density distributions $n_{\mathrm{1D}}(r,z)$ are well described within the exact thermodynamic 1D theory (see Fig.\,2a). Nevertheless, the finite coupling between the tubes leads to a small perturbation. Due to the transverse band structure the atoms slightly delocalize across the lattice which reduces the effective 1D interaction strength. This, however, can only renormalize the 1D interaction strength and the density profiles can still be described by the 1D theory. From the fit, we extract a temperature of $T\approx 35\,\pm10$\,nK in this regime for all our data sets, being comparable to the initial temperature. This confirms the adiabaticity of our loading procedure.

The situation changes however when we lower the lattice depth further and allow for a larger tunnel coupling $J$ between the tubes. We find in the experiment an increase in density relative to the 1D thermodynamic theory which appears above a critical density $n_\mathrm{crit}$ (Fig.\,2b and 2c, marked with circles). As we will discuss quantitatively later, this density increase is a clear indication of a phase transition to a 3D Bose-Einstein condensation. Qualitatively, it can be understood as follows: when the transverse coupling is strong enough the 3D density of states can be fully explored by the atoms and the atoms can condense in the ground state. The condensed atoms are maximally delocalized and reduce their interaction energy accordingly. The condensate phase is therefore more compressible which in turn translates into a density increase. Beyond this point any further increase of the tunneling coupling or the chemical potential feeds the condensate and a measurable density builds up, which goes along with long-range delocalization and phase coherence in the transverse direction. In LDA, the location where the excess density starts to build up can be converted into a critical chemical potential $\mu_{\rm crit}$.

More insight can be gained by analyzing the corresponding time of flight images. For lattice depths $s\leq20$, the TOF images show the appearance of sharp interference peaks, signaling the presence of a condensate fraction (see inset Fig.\,3). We fit a multi-peak Gaussian function to the interference pattern and determine the condensate fraction as the ratio between the number of atoms contributing to the interference peaks and the total number of atoms. As has been pointed out by several authors (see \cite{Trotzky2010} and references therein), the appearence of interference peaks is not unambiguously connected to the existence of a condensate: in the vicinity of the critical point of the 3D superfluid to Mott insulator transition less sharp interference peaks can be observed even above the critical temperature. However, the comparison between experiment and numerical simulations \cite{Trotzky2010} has revealed that for weak interactions, such as present in our study, this effect is not relevant and the interpretation as a condensate fraction is justified.

\begin{figure}
\begin{center}
\includegraphics[width=0.8\columnwidth]{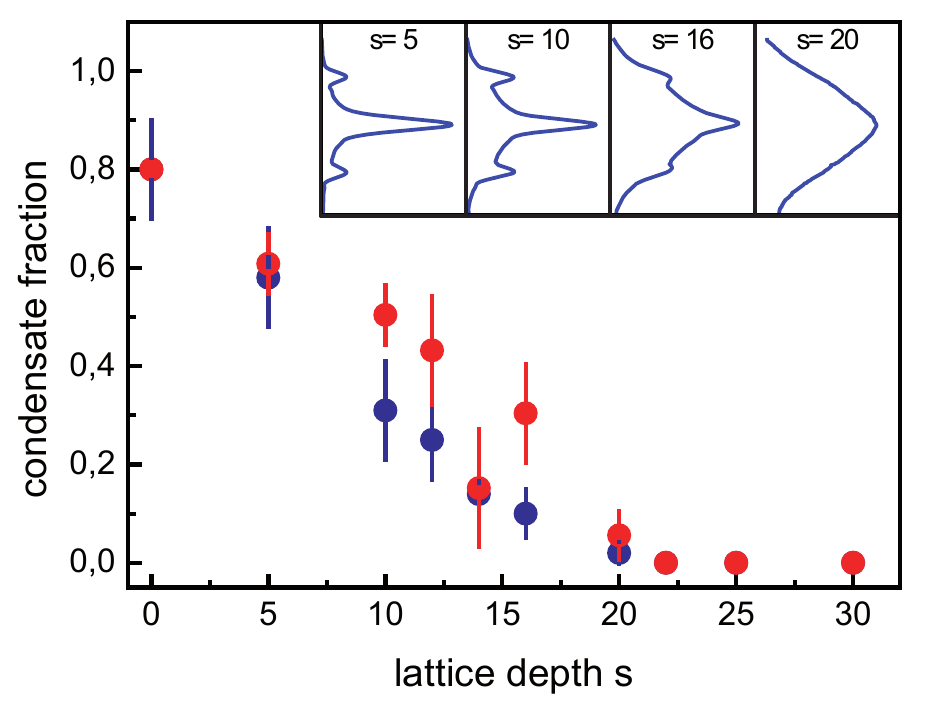}
\end{center}	
\caption{(Color online). Condensate fraction derived from TOF images (blue points, the inset shows the corresponding TOF absorption images) and excess atoms derived from the \textit{in situ} density distribution (red points). For $s>20$ no signatures of a condensate fraction or an excess density are found in the experiment.}
\label{fig:local}
\end{figure}

In order to verify whether the deviation from the 1D density profiles coincides with the formation of a condensate we determine the total number of ''excess'' atoms $N_{\mathrm {exc}}$ with $n>n_\mathrm{crit}$, and compare it with the condensate fraction found in time of flight. We first note that the central part of each line profile is well described by an inverted parabola. We therefore make a Tomas-Fermi approximation and assume a linear equation of state, $\mu=g_{\rm eff}n$, where $g_{\rm eff}$ is an effective interaction strength. For a weakly interacting condensate, the atoms fill up the trap up to the critical chemical potential. In a parabolic potential, the fraction of excess atoms to the total number of atoms is then given by $N_{\rm exc}/N_{\rm tot}=\eta(1-n_\mathrm{crit}/n_0)^{5/2}$, independent of the interaction strength. Here, $n_0$ denotes the density in the trap center. The normalization factor $\eta=0.8$ accounts for the fact that in the absence of the lattice, where $n_\mathrm{crit}/n_0$ is close to zero, we observe not more than 80 percent condensate fraction. In Fig.\,3 we compare both results. Throughout the full range of lattice depths, we find good agreement within the error bars. Hence, the total number of excess atoms is comparable the total number of condensed atoms. We therefore take the appearance of an excess density as a marker for the onset of Bose-Einstein condensation and take $\mu_{\rm crit}$ as the critical chemical potential at which the phase transition takes place. Quantum Monte Carlo simulations which were carried out on a simplified model \cite{QMC}, show that the appearance of an excess density coincides with the onset of macroscopic phase coherence. We eventually arrive at the following picture: with increasing coupling strength, a condensate fraction first develops in the center of the trap and then grows over the whole cloud. Within the condensate part, the system is three-dimensional, while in the outer parts of the cloud, the individual tubes are effectively decoupled. In this regime, the temperature determines the relevant correlation length in each tube and the influence of the transverse coupling can be neglected. The system is therefore effectively one-dimensional. Taking the system as a whole, it is a hybrid system, simultaneously hosting both dimensionalities.

\begin{figure}
\begin{center}
\includegraphics[width=\columnwidth]{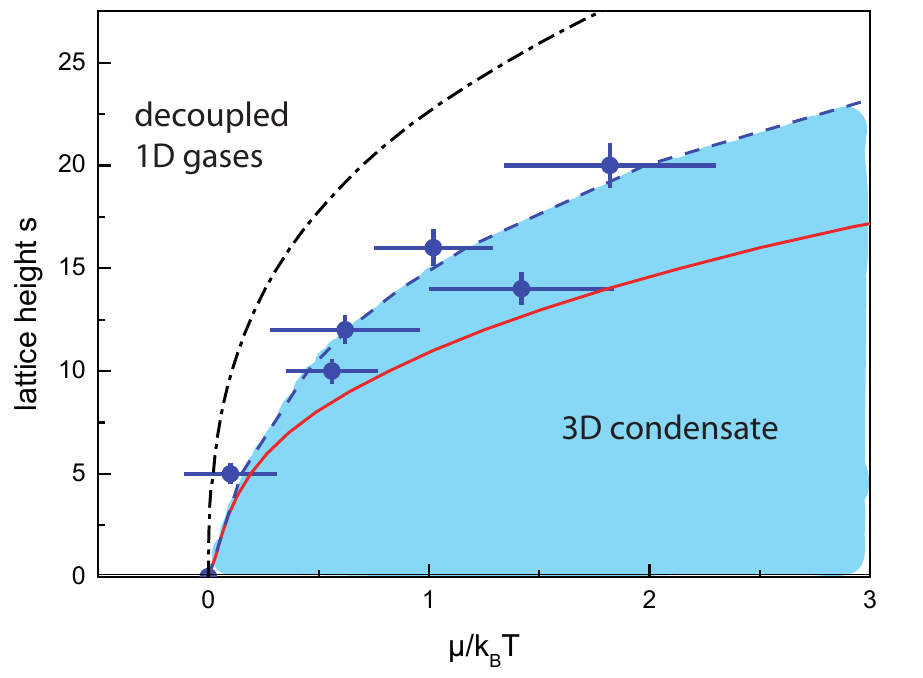}
\end{center}
\caption{(Color online). Phase-diagram in the $s-\mu$-plane. The points mark the critical chemical potential at which the phase transition takes place, the dashed line is a guide to the eye. The dashed-dotted black line is the prediction of Ref.\,\cite{Cazalilla2006}, the solid red line is the prediction of a 3D treatment of the system (see text).} 
\label{fig:phase}
\end{figure}

Analyzing the critical chemical potential for all data sets allows us to draw the phase diagram in the $s$-$\mu$ plane. This is shown in Fig.\,4. It is clearly visible that the phase transition is shifted to large positive values of the chemical potential with increasing lattice depth. The right border of the diagram displays the maximum value of the chemical potential we can reach in the experiment.

To understand the physical nature of the phase transition in more detail we compare our results with two theoretical predictions. The first one is directly taken from Ref.\ [\onlinecite{Cazalilla2006}], see Eq.~(\ref{scaling}), where the Luttinger Parameter must be determined from the density and $\mu_{\rm crit}\simeq g_{\rm 1D}(s) n_{\rm crit}$. Here, $g_{\rm 1D}(s)\simeq 2\hbar a\omega_{\perp}(s)$ is the 1D interaction strength, where $\omega_{\perp}(s)$ denotes the transverse oscillation frequency, and $a$ is the 3D scattering length. In this model, the condensation occurs because the coherence which is induced from one tube to a neighboring tube is strong enough to stabilize the condensate wave function of the global system in self-consistent way. The result is plotted as black dashed dotted line in Fig.\,4. While the trend of the experimental data is well captured by the theory, it systematically underestimates the critical chemical potential. This might have its origin in the chain mean-field treatment, which overestimates the ordered phase.
 
Alternatively, we can start from an ideal Bose gas in 3D and calculate the critical density, at which the phase transition takes place. We approximate the dispersion relation in the lattice with $E(k_z,k_x,k_y)=\hbar^2k_z^2/2m+2J(2-\cos(k_x d)-\cos(k_y d))$ and calculate the critical density by numerical integration over the Bose-Einstein distribution $N=(z^{-1}-1)^{-1}+\int Vd^3k(2\pi)^{-3}(z^{-1}\mathrm{exp}[\beta E(k_z,k_x,k_y)]-1)^{-1}$. This corresponds to the standard picture that condensation occurs when the population in the excited states is saturated. We restrict our calculation to the lowest band. Without a lattice, the interaction between the atoms leads only to a small shift of the phase transition to positive values of the chemical potential. But now, the transverse confinement in the tubes has a two-fold influence. On the one hand, the 1D interaction strength within each tube increases with the confinement. On the other hand, the transverse band structure leads to a higher effective mass and a shortening of the thermal deBroglie wavelength in this direction. This results in a higher density necessary for achieving the condition for Bose-Einstein condensation. For the resulting critical chemical potential ($\mu_{\rm crit}\simeq g_\mathrm{1D}(s)n_\mathrm{crit}$), both effects magnify each other and the phase transition is shifted significantly. The prediction is plotted as solid red line in Fig.\,4. Despite the approximative character of the model, it describes the data for small values of $s$ rather well, while for larger values of $s$ it overestimates the critical chemical potential.

\begin{figure}
\begin{center}
\includegraphics[width=0.9\columnwidth]{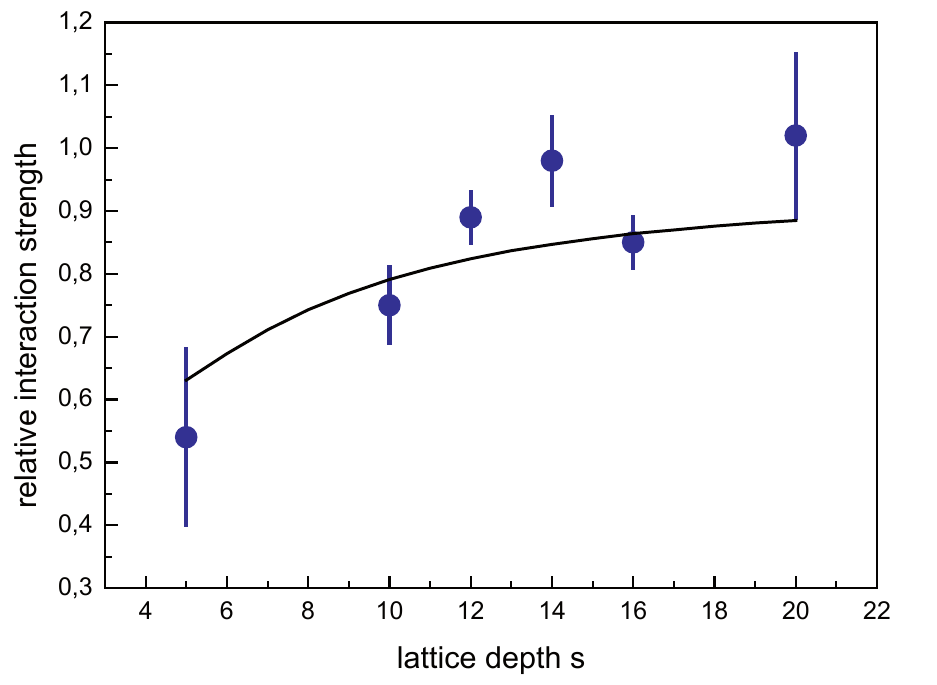}
\end{center}
\caption{(Color online). Ratio between the effective interaction for the coupled system and that for an isolated tube. Blue points are the experimental data and the black curve is the theoretical prediction (see text).} 
\label{fig:int-strength}
\end{figure}

Within the condensate phase, no exact solution is known at finite temperature. We can now use our data to extract the effective interaction strength for our range of parameters, where the chemical potential is comparable to the thermal energy. To this end, we fit the condensate part of the density with a Thomas Fermi profile and compare it to a Thomas Fermi fit of the 1D thermodynamic theory. In Fig.\,5 we show the ratio between the effective interaction strength and the corresponding interaction strength of the isolated 1D system. For large tunnel coupling the effective interaction is reduced by almost a factor of 2 compared to the uncoupled system, while for decreasing coupling strength, the difference gets smaller. This has its origin in the delocalization of the atoms in the lowest Bloch state which is absent for an isolated tube. We can give a rough estimation of this reduction by integrating out the two transverse directions for the lowest Bloch state and compare the resultiong interaction strenght to the result of an isolated tube, $g_{\rm 1D}(s)\simeq 2\hbar a\omega_{\perp}(s)$. The ratio between the two is shown as black curve in Fig.\,5 and follows the trend of the data. This exemplifies how a cross-dimensional phase transition combines properties from both dimensionalities: While the system above $\mu_{\rm crit}$ is a 3D condensate, the effective interaction strength of this phase is partially inherited from the underlying 1D geometry. 

Our experiments reveal the coherence properties, the phase diagram and the effective interaction strength of coupled one-dimensional quantum systems. The crossover from one-dimensional to higher dimensional behavior is a universal phenomenon, which can be observed in a plethora of coupled low-dimensional systems, such as spin ladders \cite{Tsvelik1999,Ruegg2008,Thielemann2009}, spin chains \cite{Affleck2002,chains} and quantum wires \cite{Seroussi2014}, 
which are all described by quasi-one-dimensional interacting effective boson systems. In these systems a systematic analysis of the phase transition and the order parameter as a function of transverse coupling $J$ relied so far on mean field theoretical arguments. The ability to perform thermodynamic studies on coupled 1D ultracold gases now allows to map out the phase diagram as a function of coupling strength and chemical potential. The theoretical models provide a corridor for the experimental data, demonstrating qualitative agreement, but at the same time reveals the need for further investigation. Our results can help to benchmark numerical simulations of coupled one-dimensional system, providing also a better description of real materials.

\begin{acknowledgments}
We are thankful for useful discussion with Axel Pelster and Tim Langen. We thank D. Morath and D. Stra{\ss}el for providing results of QMC calculations. We acknowledge financial support by the DFG within the SFB/TRR 49, and the MAINZ graduate school. V.G. and G.B. are supported by Marie Curie Intra-European Fellowships. 
\end{acknowledgments}

\bibliographystyle{apsrev4-1}


\bibliography{0_Literature_Andreas}

\end{document}